\documentclass[aps,showpacs,preprint,preprintnumber,nofootinbib,amsmath,amssymb,ascmac,bm,12pt]{revtex4}
\usepackage{bm}
\usepackage[dvipdfm]{graphicx}
\begin{document}
\title{\vspace*{0.5cm}
Topology Changing Process of Coalescing Black Holes on Eguchi-Hanson Space
}
\author{%
${}^{1}$Masashi Kimura\footnote{E-mail:mkimura@sci.osaka-cu.ac.jp},~%
${}^{1}$Hideki Ishihara\footnote{E-mail:ishihara@sci.osaka-cu.ac.jp},~%
${}^{2}$Shinya Tomizawa\footnote{E-mail:tomizawa@post.kek.jp}\\%
~and~
${}^{3}$Chul-Moon Yoo\footnote{E-mail:c\_m\_yoo@apctp.org}%
}
\affiliation{
${}^{1}$Department of Mathematics and Physics,\\
Graduate School of Science, Osaka City University,\\
3-3-138 Sugimoto, Sumiyoshi, Osaka 558-8585, Japan
\vspace{0.2cm}\\
${}^{2}$Cosmophysics Group, Institute of Particle and Nuclear Studies, \\
KEK, Tsukuba, Ibaraki, 305-0801, Japan
\vspace{0.2cm}\\
${}^{3}$Asia Pacific Center for Theoretical Physics,\\
Pohang University of Science and Technology, Pohang 790-784,~Korea
}
\begin{abstract}
We numerically study the event horizons
of two kinds of five-dimensional coalescing black hole solutions with different asymptotic
structures: the five-dimensional Kastor-Traschen solution (5DKT) and the coalescing black hole solution
on Eguchi-Hanson space (CBEH).
Topologies of the spatial infinity are ${\rm S}^3$ and $L(2;1)={\rm S}^3/{\mathbb Z}_2$, respectively.
We show that the crease sets of event horizons are topologically ${\rm R}^1$ in 5DKT and
${\rm R}^1\times {\rm S}^1$ in CBEH, respectively.
If we choose the time slices
which respect space-time symmetry,
the first contact points of
the coalescing process is a point in the 5DKT case but a ${\rm S}^1$ in the CBEH case.
We also find that
in CBEH, time slices can be chosen so that a black ring with ${\rm S}^1\times {\rm S}^2$
topology can be also formed during a certain intermediate period unlike the 5DKT.
\end{abstract}

\preprint{OCU-PHYS 315}
\preprint{AP-GR 67}
\preprint{APCTP Pre2009 - 007}
\preprint{KEK-TH 1320}


\date{\today}
\maketitle

\section{Introduction}
One of the most interesting predictions of string theory,
which is a strong candidate of the unified theory, 
is that our world is a higher dimensional space-time.
Since we feel we live in the four-dimensional space-time in low energy physical phenomena, 
extra-dimensions should be effectively compactified. 
It is natural to consider that the space-time is locally a direct product of our
four-dimensional space-time and extra-dimensions.
However, there are many possibilities how four-dimensional space-time and
extra-dimensions are connected globally.
The structure of the whole higher-dimensional space-time is not known.

Can we get any information of the global structure of
the whole higher-dimensional space-time from
any experiment localized in a finite region?
To approach this question, at the first step,
we consider a system of black holes with a non-trivial
asymptotic structure as a toy model and
we compare the coalescing process with a trivial asymptotic structure case.

In the series of works~\cite{Ishihara:2006ig,Yoo:2007mq,Matsuno:2007ts}, 
it is presented that differences of black hole coalescence processes caused 
by asymptotic structure.
In the case of the five-dimensional
space-time in which the boundary of the spatial infinity
is not ${\rm S}^3$ but lens space,
it is possible that two black
holes with the topology of ${\rm S}^3$ coalesce and 
change into a single black hole with the topology
of the lens space,
i.e., ${\rm S}^3 + {\rm S}^3 \to L(2;1) = {\rm S}^3/{\mathbb Z}_2$.
Analyzing the horizons at early time and late time, we found that
the areas of the horizon after coalescence in such
cases is larger than the ordinary cases in five dimension, i.e., ${\rm S}^3 + {\rm S}^3 \to {\rm S}^3$.
However, the shape of the event horizon is not clear in an intermediate period.
In this paper, 
we discuss details of event horizon structures in
coalescing black holes with non-trivial asymptotic structures.\footnote{
Recently, the structure of the event horizons of lens space
was also discussed in~\cite{Ida:2009nd}.}

The paper is organized as follows: 
in Section II we briefly review the coalescing black holes with 
trivial and non-trivial asymptotic structures in five dimensions.
Section III we investigate the structures of the event horizon of coalescing black holes 
numerically and discuss coalescing process in typical time slices.
Section IV is devoted to the discussion of the structure of the crease set of the event horizon.
Finally, in Section V we discuss the obtained results.

\section{Coalescing Black Holes in Five Dimensions}
Black hole coalescence is one of the most interesting issues in a study of gravity.
To treat the coalescencing process,
we need heavy numerical work in general. 
However, if the mass and the electric charge of each black hole are equal, 
we can construct exact solutions
which describe the coalescencing processes driven
by a positive cosmological constant~\cite{Kastor:1992nn,London:1995ib,
Brill:1993tm,Nakao:1994mm,Ida:1998qt,
Ishihara:2006ig,Matsuno:2007ts,Kimura:2009er}.\footnote{
If we set the cosmological constant to zero, the solution~\cite{Kastor:1992nn,London:1995ib}
reduces to the Majumdar-Papapetrou solution~\cite{Majumdar:1947eu,Papaetrou:1947ib,Hartle:1972ya} 
which describes static multi-black holes.
The higher-dimensional generalizations of the multi-black holes are discussed in
\cite{Myers:1986rx,Gauntlett:2002nw,Ishihara:2006iv,Ishihara:2006pb}, and recently
the smoothness of horizons of higher-dimensional multi-black holes are also investigated
in \cite{Welch:1995dh,Candlish:2007fh,Candlish:2009vy,Kimura:2008cq}.
}

One of such solutions in five-dimensional Einstein-Maxwell theory with a positive cosmological constant is the 
Kastor-Traschen solution(5DKT)~\cite{Kastor:1992nn,London:1995ib}. 
The metric and gauge 1-form for 5DKT are given by
\begin{eqnarray}
ds^2 &=& -H^{-2} dt^2 + H e^{-\lambda t} \left[ dx^2 + dy^2 + dz^2 + dw^2 \right],
\label{5dktmet}
\\
A &=& \pm \frac{\sqrt{3}}{2}H^{-1}dt,
\end{eqnarray}
with
\begin{eqnarray}
H &=& 1+\frac{1}{e^{-\lambda t}} \left(\frac{m_1}{x^2+y^2+z^2+(w- a)^2} + \frac{m_2}{x^2+y^2+z^2+(w+ a)^2}\right),
\end{eqnarray}
respectively, 
where $\lambda = 2\sqrt{\Lambda/3}$ and $\Lambda$ is a positive cosmological constant.
This solution describes the physical process such that
two black holes with the topology of ${S}^3$
coalesce into a single black hole  with the topology of ${ S}^3$.

Recently, 
coalescing black hole solution on Eguchi-Hanson space(CBEH)~\cite{Ishihara:2006ig} 
has been found as another exact solution in the same theory.
The metric and gauge 1-form for this solution are given by
\begin{eqnarray}
ds^2 &=& -H^{-2} dt^2 + H e^{-\lambda t} \left[ V^{-1} (dx^2 + dy^2 + dz^2) + V ((a/8)d\psi + \omega)^2 \right],
\label{eq:EH}
\\
A &=& \pm \frac{\sqrt{3}}{2}H^{-1}dt,
\end{eqnarray}
with
\begin{eqnarray}
H &=& 1+\frac{1}{e^{-\lambda t}}\left(\frac{M_1}{\sqrt{x^2+y^2+(z- a)^2}}+\frac{M_2}{\sqrt{x^2+y^2+(z+ a)^2}}\right),
\\
V^{-1} &=&\frac{a/8}{\sqrt{x^2+y^2+(z- a)^2}}+\frac{a/8}{\sqrt{x^2+y^2+(z+ a)^2}},
\\
\omega
&=&
\frac{a}{8}
\left[\frac{z-a}
{\sqrt{x^2+y^2+(z-a)^2}}
+
\frac{z + a}
{\sqrt{x^2+y^2+(z+a)^2}}
\right]\frac{xdy-ydx}{x^2+y^2},
\label{eq:EH8}
\end{eqnarray}
where we note that the metric inside the square bracket in (\ref{eq:EH}) 
is the metric of the Eguchi-Hanson space~\cite{Eguchi:1978xp,Gibbons:1979zt} 
which has a non-trivial asymptotic structure called the lens space $L(2;1) = {\rm S}^3/{\mathbb Z}_2$.

This solution
describes the physical process such that two black holes with the topology
of ${\rm S}^3$ coalesce into a single black hole with the topology of the lens space
$L(2;1)$ due to the non-trivial asymptotic structure~\cite{Ishihara:2006ig}.
To confirm this, let us see the behavior of the metric at early time $t \to -\infty$
and at late time $t \to \infty$, following the discussion in~\cite{Ishihara:2006ig}.
At early time $t\to -\infty$ and $R_{i}\to 0$, the metric behaves as
\begin{eqnarray}
ds^2 &\simeq& -\left(1 + \frac{m_i}{e^{-\lambda t}r_i^2}\right)^{-2} dt^2 
\notag               \\ 
                &&+ \left(1 + \frac{m_i}{e^{-\lambda t} r_i^2}\right)e^{-\lambda t}
                                  \left[
                                    dr_i^2 + \frac{r_i^2}{4}\Big(d\theta^2 +  \sin^2\theta d\phi^2 
                                                         + \left(d\psi + \cos \theta d\phi  \right)^2
                                                         \Big)
                                 \right],
\label{earlymet}
\end{eqnarray}
where we have introduced a new radial coordinate, $r_i^2:=R_i a/2$ and a new mass parameter, $m_i:=M_ia/2$.
This metric is the same form as that of five-dimensional
Reissner-Nordstr\"om-de Sitter solution with mass parameter $m_i$ written in the cosmological coordinate.
Therefore, we can see that there are two black holes with the topology of ${\rm S}^3$ at early time.
On the other hand, 
at late time $t\to \infty$ and $R \to \infty$,
the metric behaves as
\begin{eqnarray}
ds^2 &\simeq & -\left(1 + \frac{2(m_1+m_2)}{r^2}\right)^{-2} dt^2 
\notag               \\ 
                &+& \left(1 + \frac{2(m_1+m_2)}{e^{-\lambda t}r^2}\right)
                                e^{-\lambda t}  \left[
                                    dr^2 + \frac{r^2}{4}\left(d\theta^2 +  \sin^2\theta d\phi^2 
                                                         + \left(\frac{d\psi}{2} + \cos \theta d\phi  \right)^2
                                                         \right)
                                 \right],
\label{latemet}
\end{eqnarray}
where we introduce the new radial coordinate $r^2:=a R$.
This resembles the metric of the five-dimensional Reissner-Nordstr\"om-de Sitter solution
with mass equal to $2(m_1 + m_2)$ but the topology of horizon is the lens space $L(2;1)$.
Therefore, we can see the solution (\ref{eq:EH}) describes the process such that
two black holes with ${\rm S}^3$ at early time coalesce into a single black hole with the lens space $L(2;1)$ 
at late time.

However, since 
here, we only investigate the asymptotic behaviors of the metric at early
time and at late time,
it is not clarified how two black holes with ${\rm S}^3$ coalesce into a single black hole with
the lens space $L(2;1)$.
So, in the following sections, we numerically investigate the location and the shape of
the event horizon for the solution (\ref{eq:EH})-(\ref{eq:EH8})
and make clear the process of the coalescence.

\section{Event Horizons}
\begin{figure}[!t]
\begin{center}
\includegraphics[width=0.65\linewidth,clip]{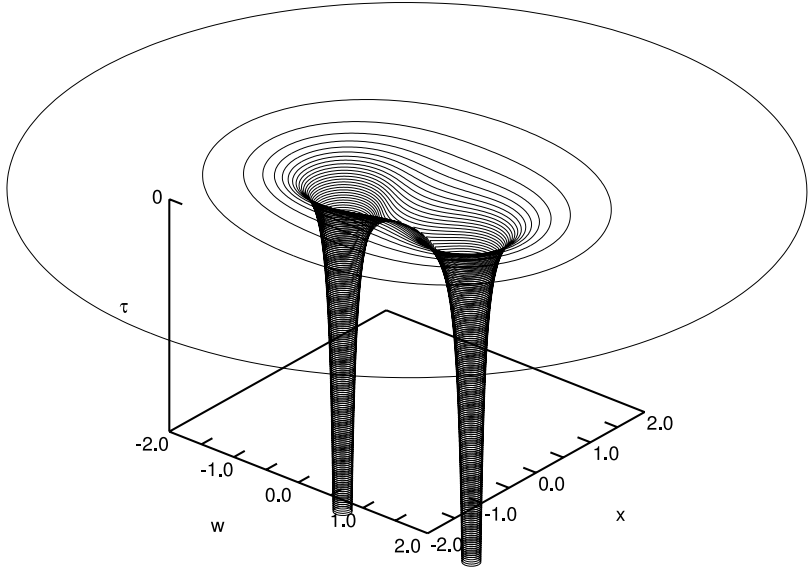}
\end{center}
\caption{Event horizon of five dimensional Kastor-Traschen solution.
         ($m_1=m_2=1, a=1, \lambda =1/(2\sqrt{2})$)}\label{fig:5DKT}
\end{figure}
\begin{figure}[!t]
\begin{center}
\includegraphics[width=0.45\linewidth,clip]{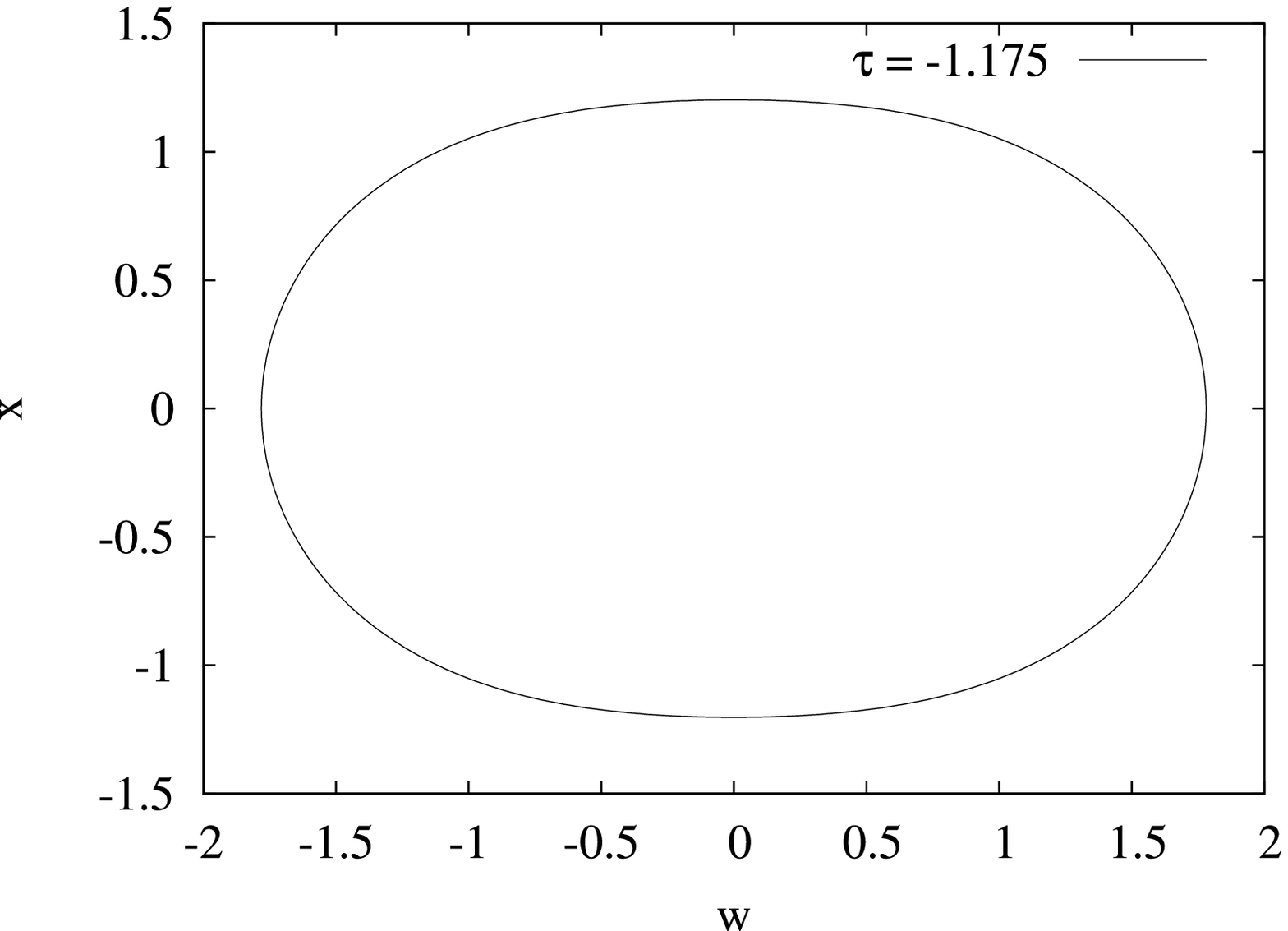}
\includegraphics[width=0.45\linewidth,clip]{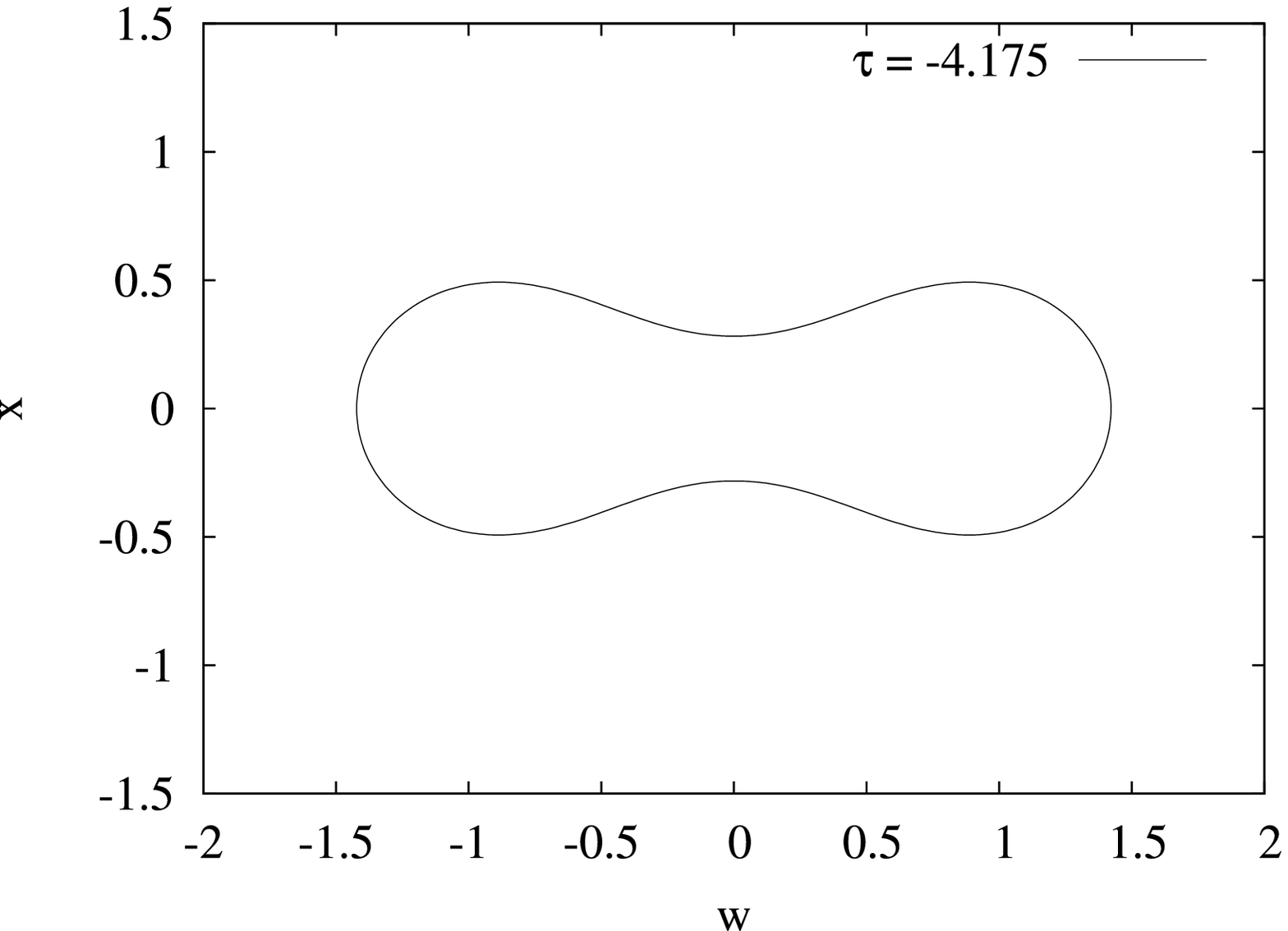}
\includegraphics[width=0.45\linewidth,clip]{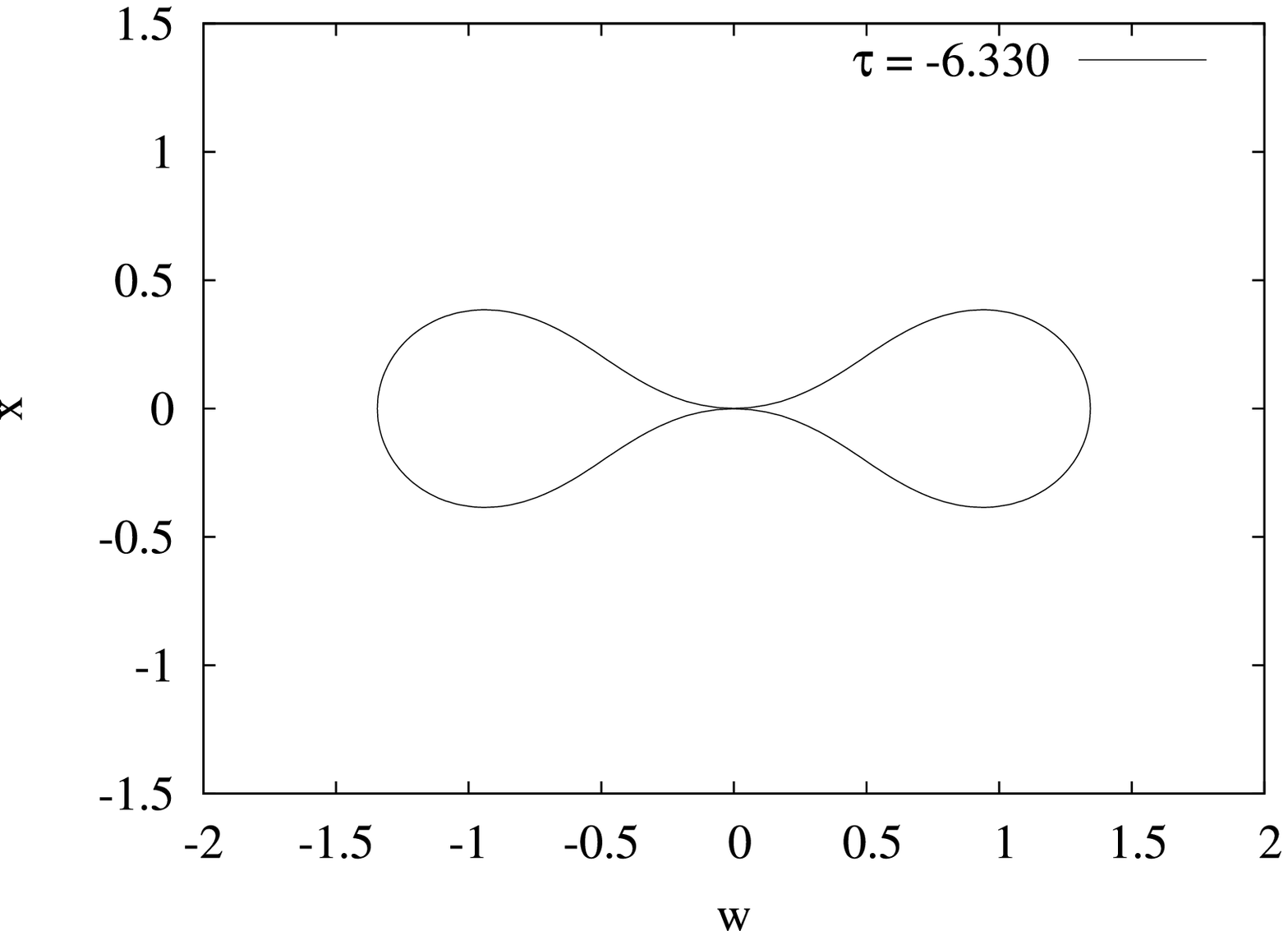}
\includegraphics[width=0.45\linewidth,clip]{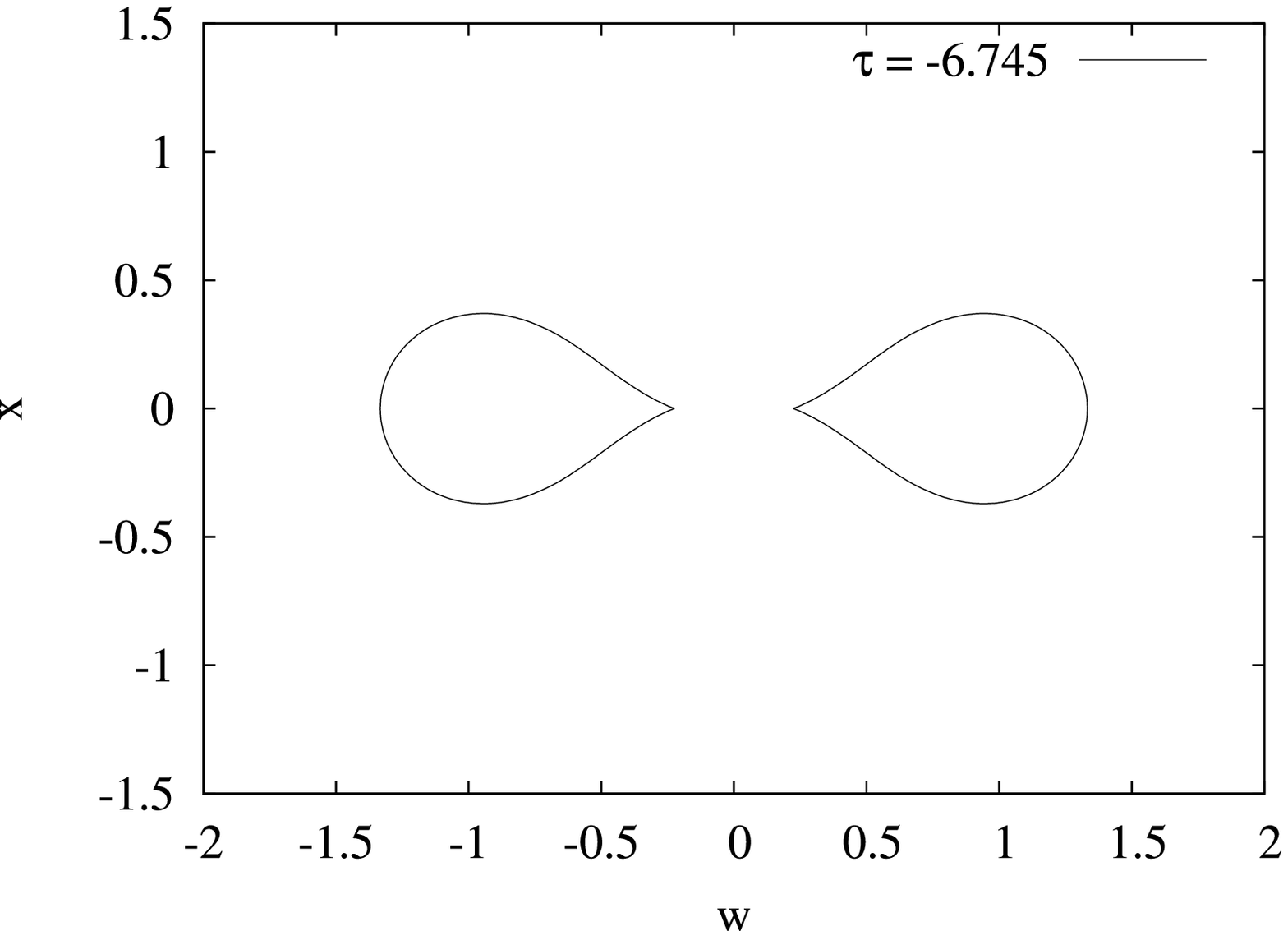}
\includegraphics[width=0.45\linewidth,clip]{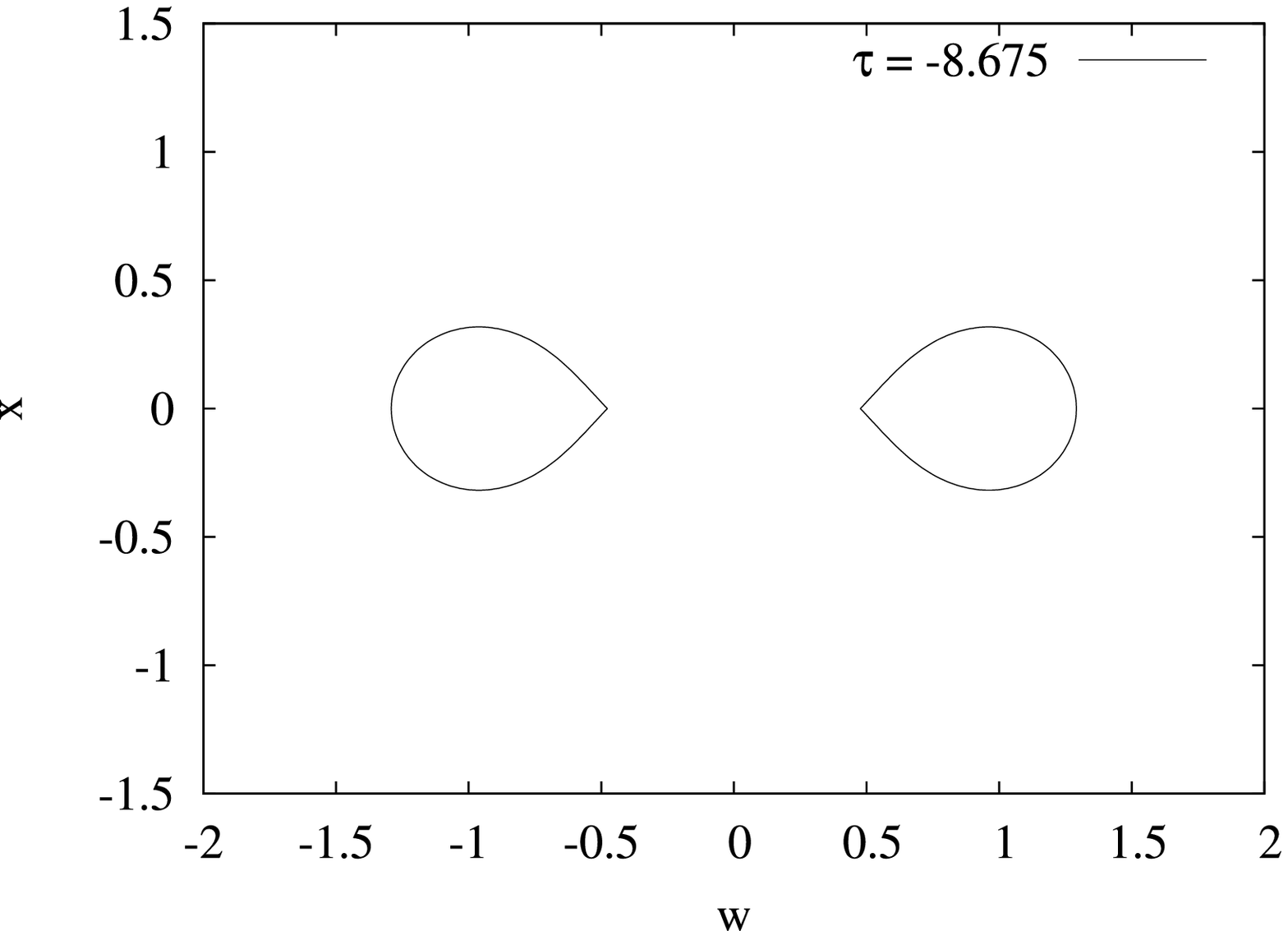}
\includegraphics[width=0.45\linewidth,clip]{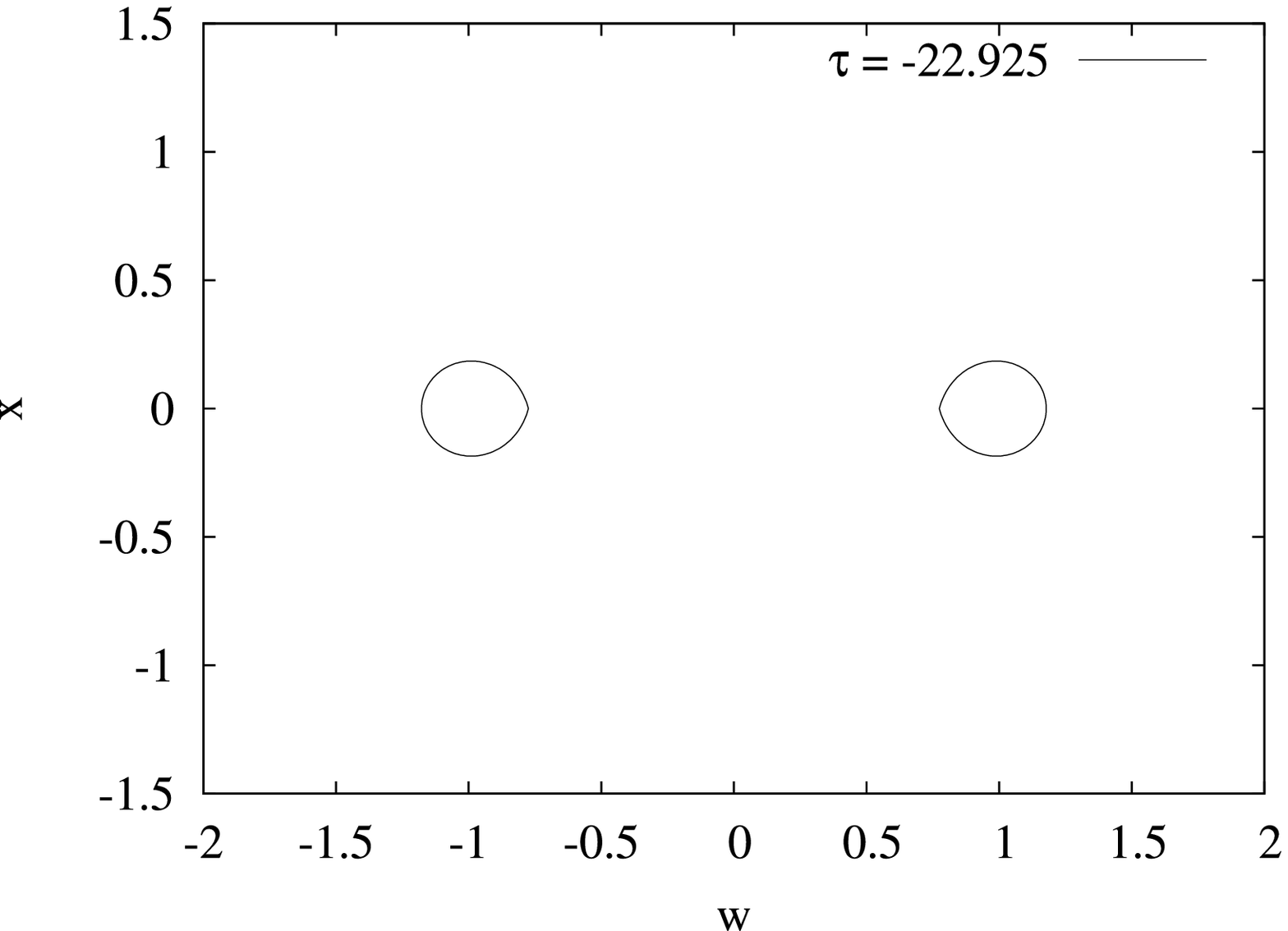}
\end{center}
\caption{Coordinate value of event horizon of each time slices in five dimensional Kastor-Traschen solution.}
\label{fig:5DKTtimeslice}
\end{figure}
\begin{figure}[!t]
\begin{center}
\includegraphics[width=0.65\linewidth,clip]{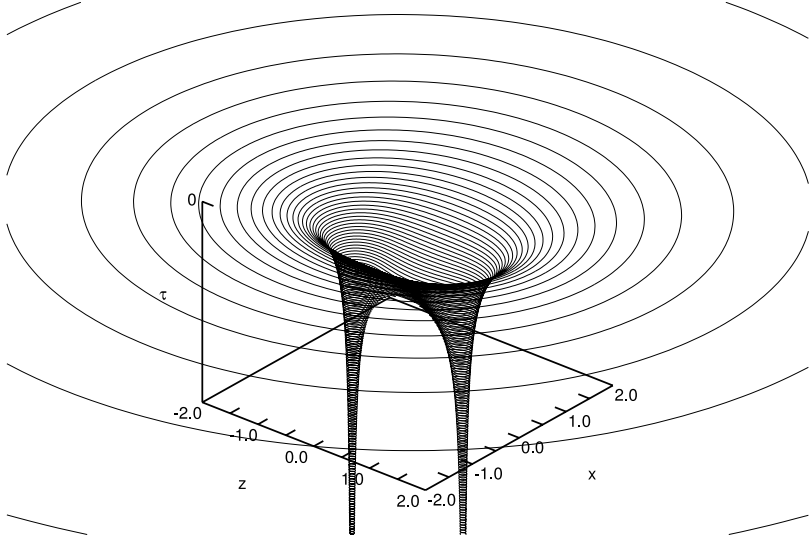}
\end{center}
\caption{Event horizon of coalescing black holes on Eguchi-Hanson space.
         ($M_1=M_2=2, a=1, \lambda =1/(2\sqrt{2})$)}\label{fig:EH}
\end{figure}
\begin{figure}[!t]
\begin{center}
\includegraphics[width=0.45\linewidth,clip]{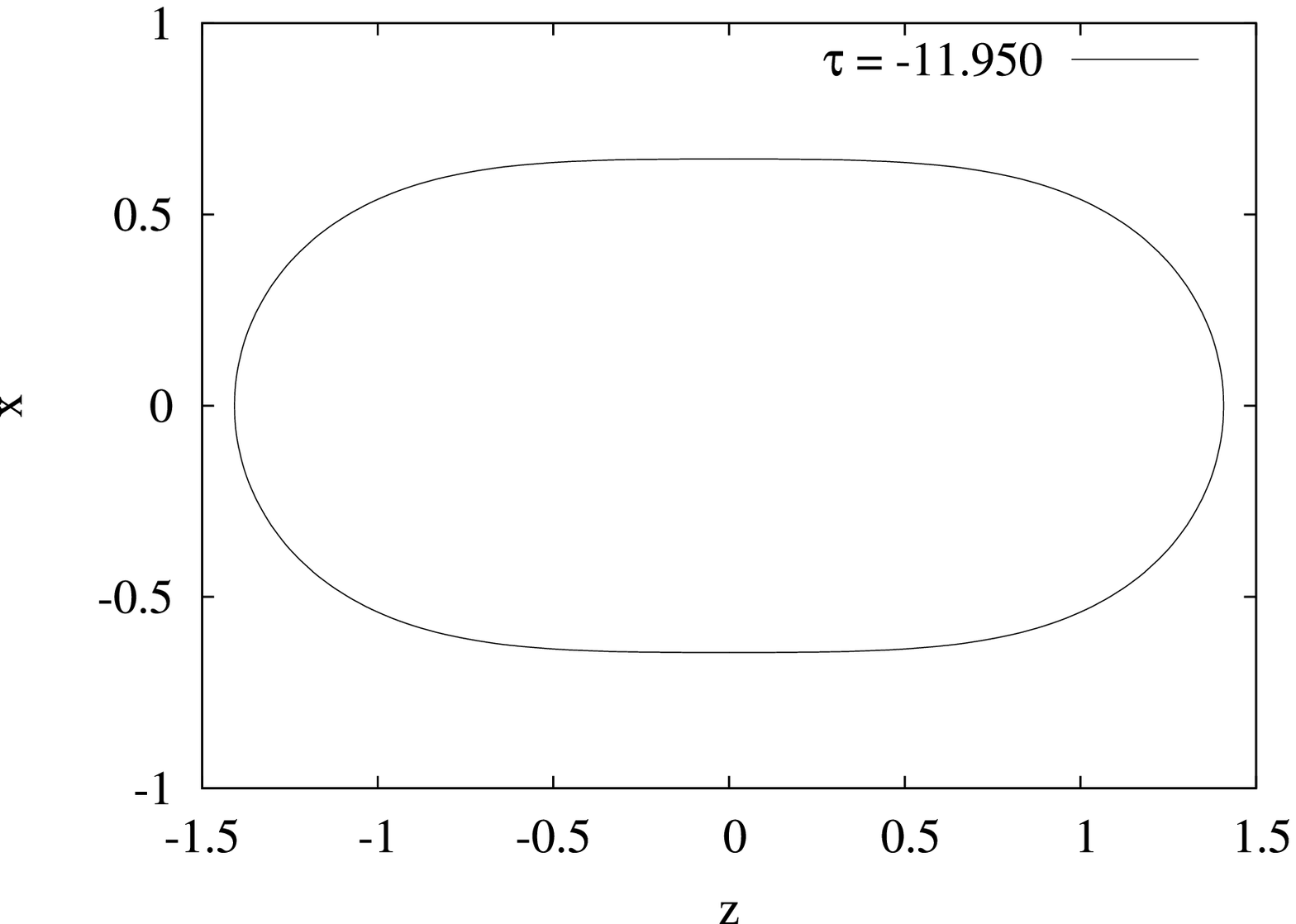}
\includegraphics[width=0.45\linewidth,clip]{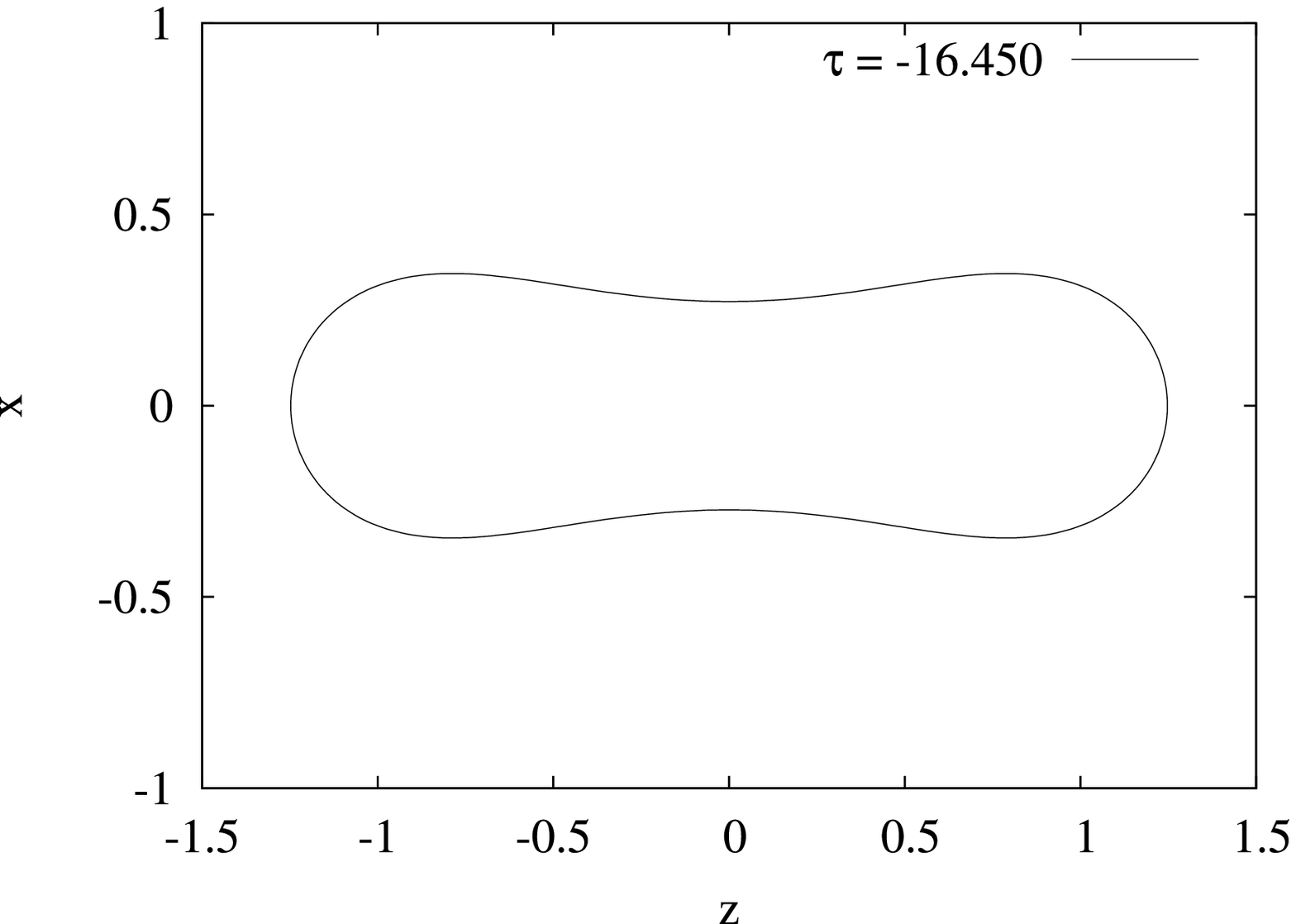}
\includegraphics[width=0.45\linewidth,clip]{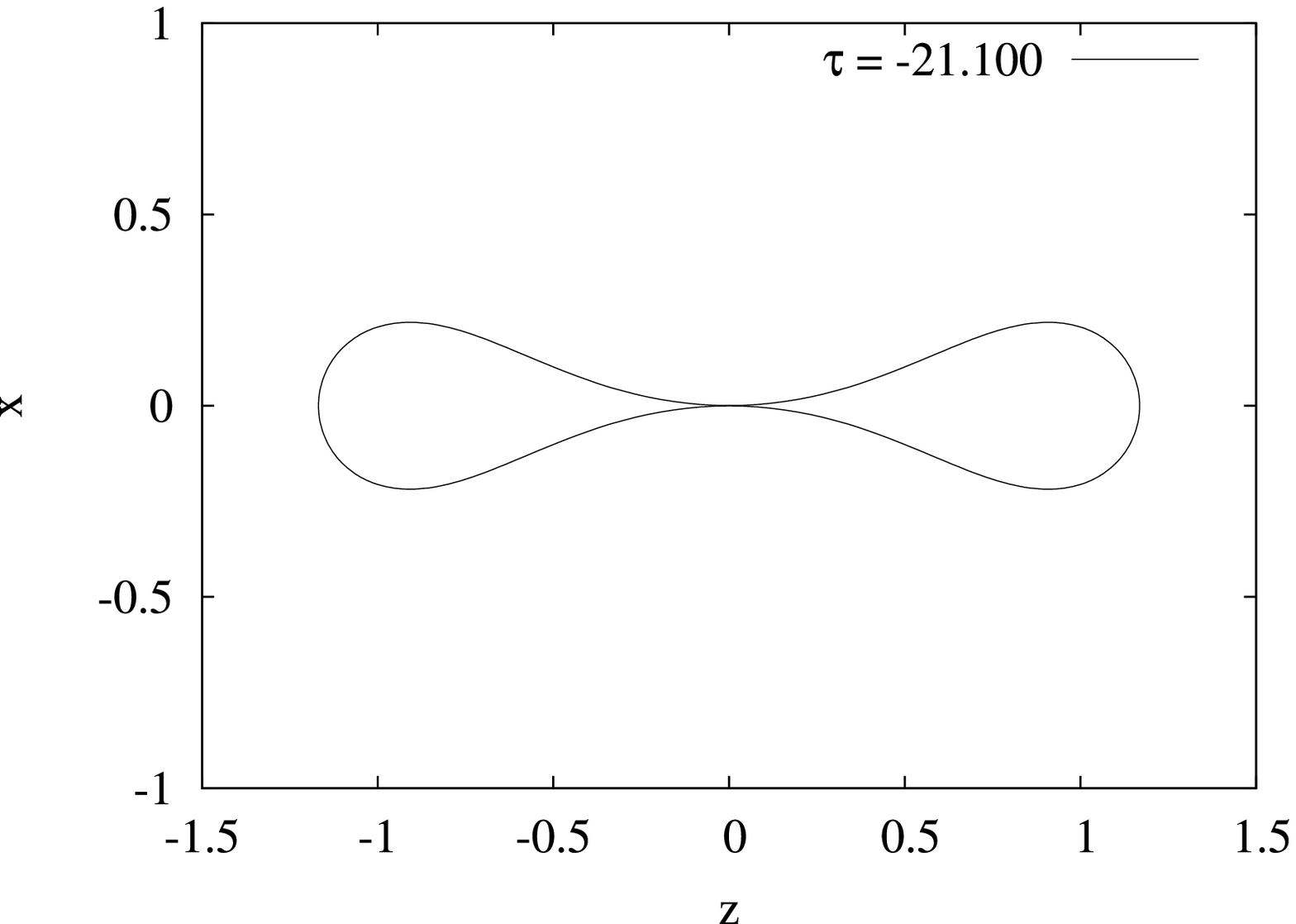}
\includegraphics[width=0.45\linewidth,clip]{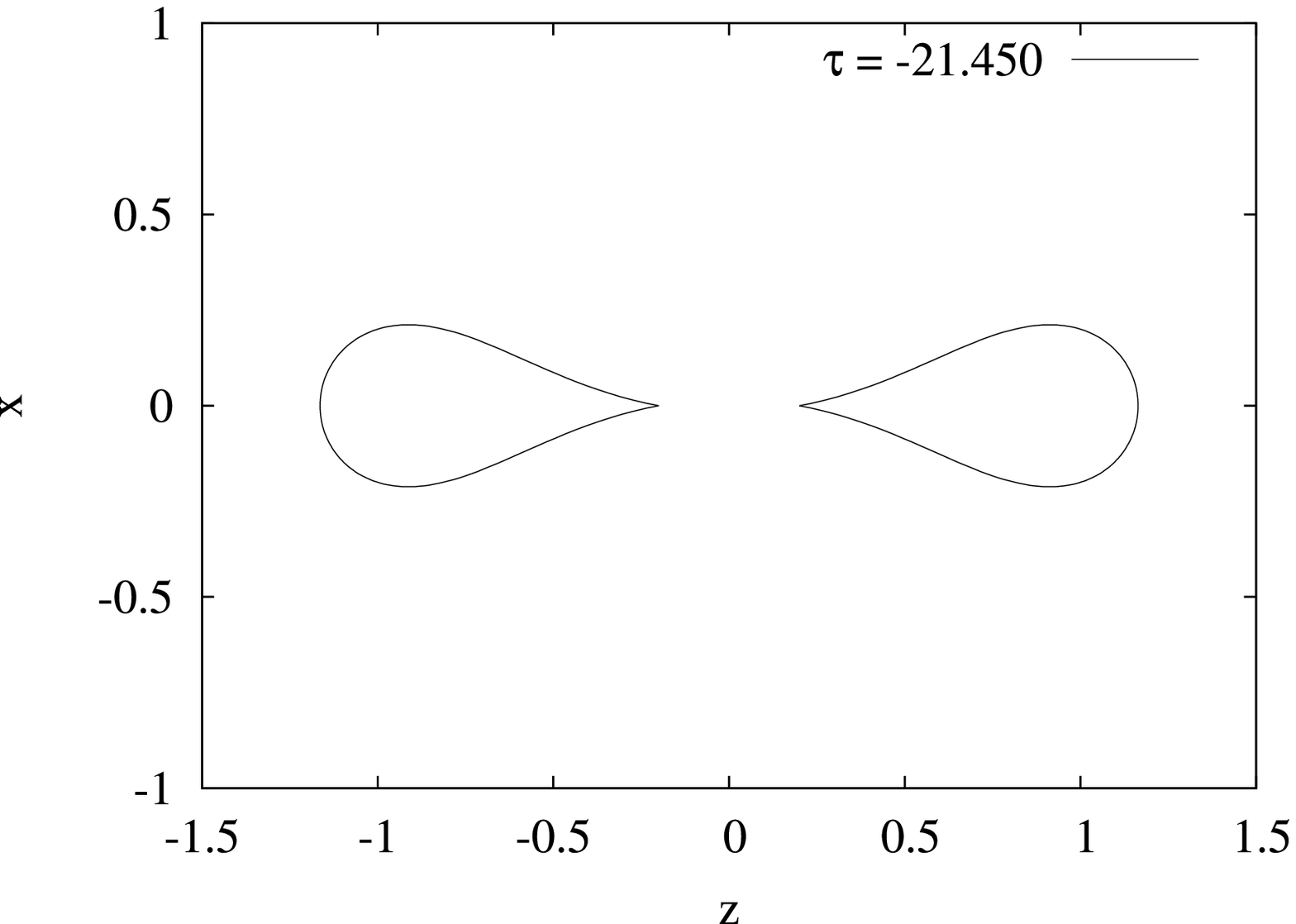}
\includegraphics[width=0.45\linewidth,clip]{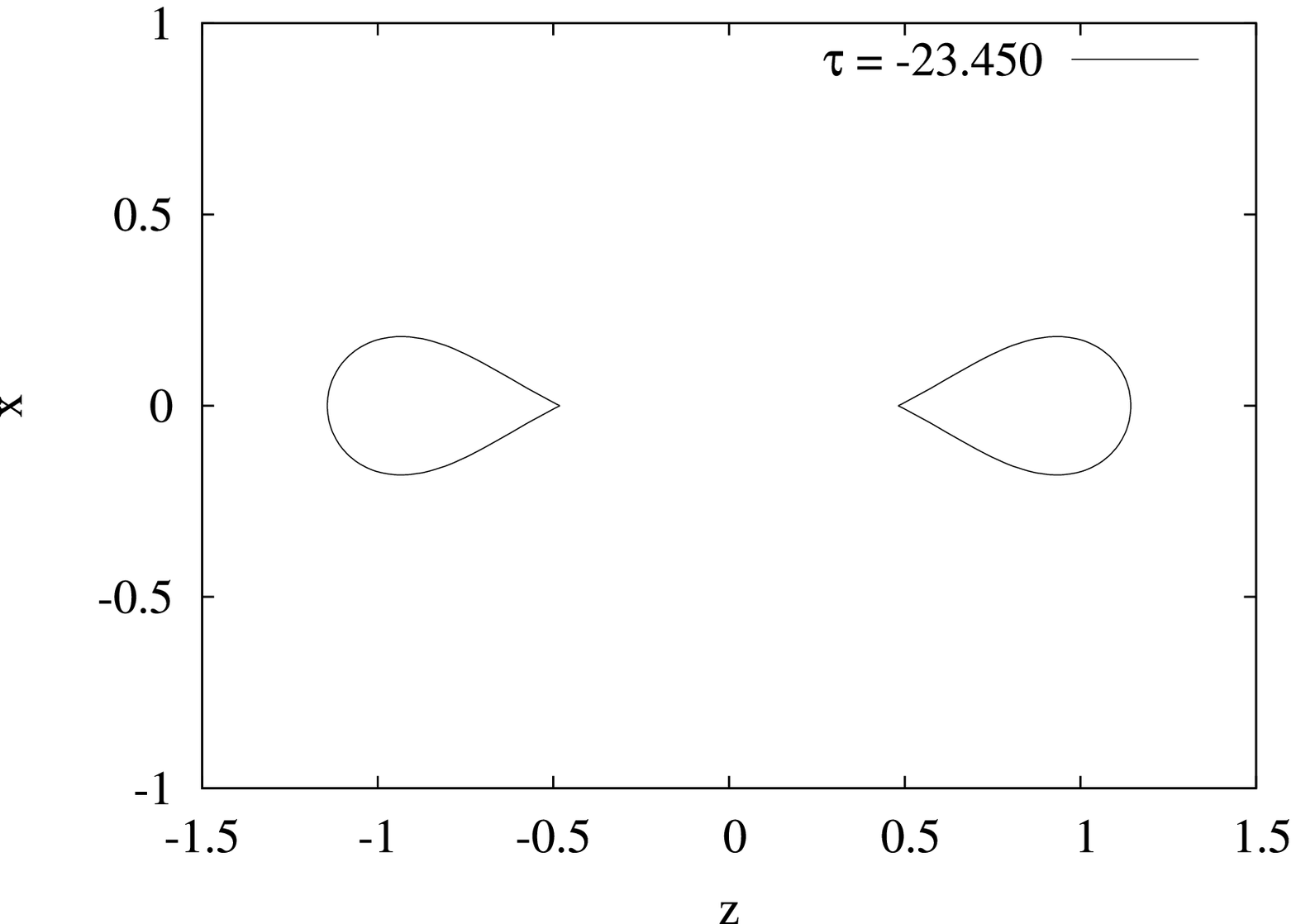}
\includegraphics[width=0.45\linewidth,clip]{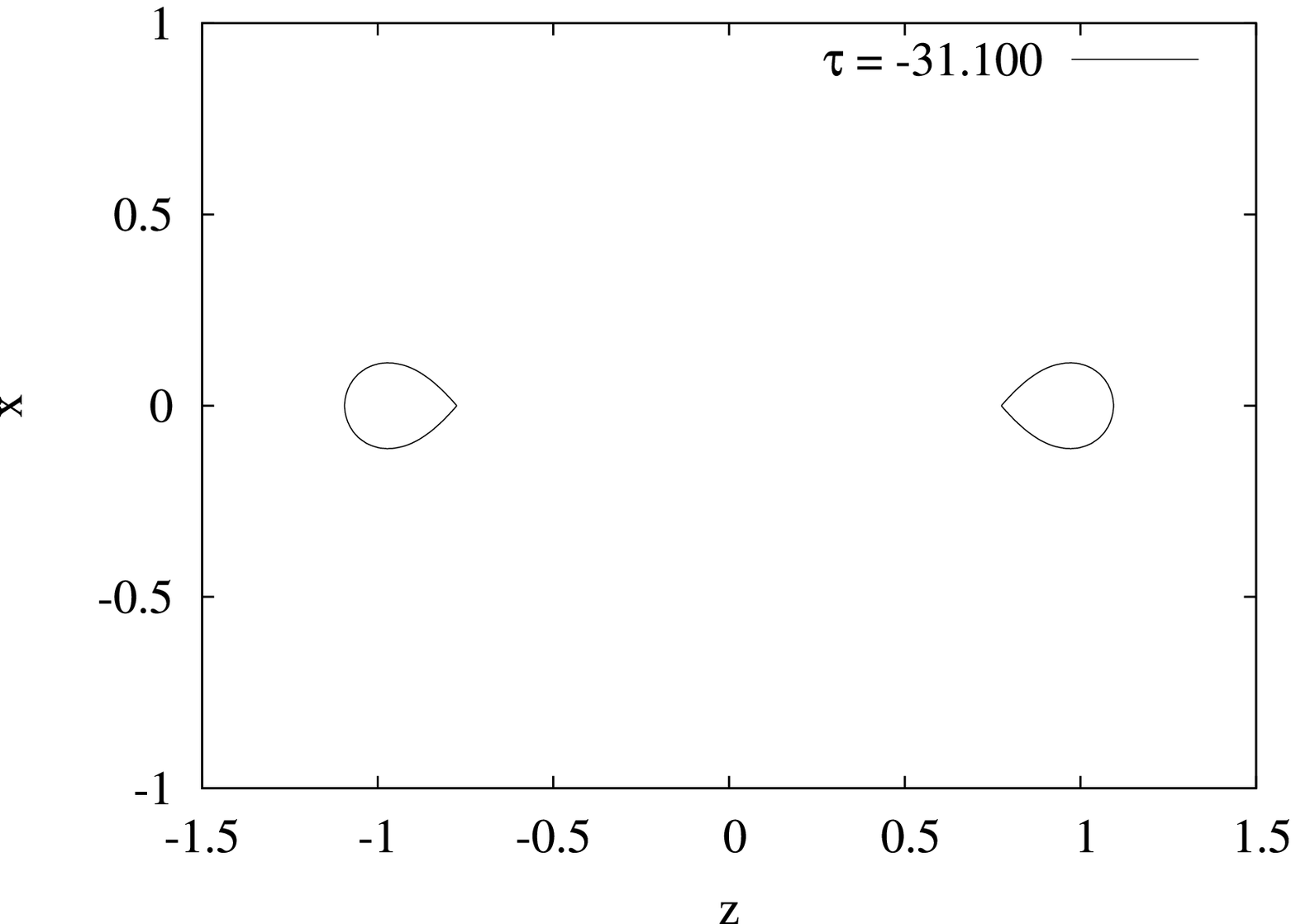}
\end{center}
\caption{Coordinate value of event horizon of each time slices in coalescing black holes on Eguchi-Hanson space.}
\label{fig:EHtimeslice}
\end{figure}
An event horizon is defined as the boundary of the causal past of the future null infinity. 
Due to the absence of Killing horizons in a dynamical spacetime, it is difficult to determine
the location of an event horizon analytically.
We can investigate  the location of the event horizon 
by integrating numerically null geodesic equations backward from sufficiently future to the past\cite{Ida:1998qt}.

Since both metrics Eqs.(\ref{5dktmet}) and (\ref{eq:EH}) asymptote to the metric of 
Reissner-Nordstr\"om-de Sitter solution in the limit of $t\to \infty$, fortunately we can know
where the event horizons locate at late time.
So if we solve null geodesic equations numerically from each point of the spatial cross section
of the event horizons at late time backward the past where we set the initial null direction
to be tangent to the event horizon at late time,
we can get null geodesic generators of the event horizons,
namely we can find the locations of the event horizons.

\subsection{5DKT case}
In 5DKT case, we obtain null geodesic generators of the event horizon for the metric (\ref{5dktmet}).
We can set $y=0,z=0$ without loss of generality because of the $SO(3)$ symmetry in $(x,y,z)$ space.
Then, we solve null geodesics in the effective three-dimensional metric given by
\begin{eqnarray}
ds^2 &=&
 -H^{-2}dt^2
+ H e^{-\lambda t} \left[ dx^2  + dw^2 \right],
\\
H &=& 1+\frac{1}{e^{-\lambda t}} \left(\frac{m_1}{x^2+(w- a)^2} + \frac{m_2}{x^2+(w+ a)^2}\right).
\end{eqnarray}
After some numerical calculations,
we get the data of coordinate values of location of event horizon,
and we plot the coordinate values
$\tau:= e^{-\lambda t},~x,~w$ in Fig.\ref{fig:5DKT}.
In this graph, each contour line means 
cross section of horizon with a
$\tau = \rm{const}$. surface.
In Fig.\ref{fig:5DKTtimeslice},
we plot the the coordinate values
$x,~w$ of event horizon at some typical time slices.
{}From Fig.\ref{fig:5DKTtimeslice} we can see 
that two black holes collide at $\tau = -6.330$
and the first contact point
is given by $x=0, w=0$.
Since $w$-axis is fixed points $SO(3)$ symmetry in $(x,y,z)$ space,
the topology of this first contact point in Fig.\ref{fig:5DKTtimeslice}
is a point 
in the whole five-dimensional space-time.
\subsection{CBEH case}
In CBEH case, we obtain null geodesic generators of the event horizon for the metric (\ref{eq:EH}).
We can set $y=0$ without loss of generality because of the $SO(2)$ symmetry in $(x,y)$ space.
Furthermore, since $\partial_\psi$ is a Killing vector, we can set $\psi = 0$.
Then, we solve null geodesics in the effective three-dimensional metric given by
\begin{eqnarray}
ds^2 &=& -H^{-2} dt^2 + H e^{-\lambda t} V^{-1} (dx^2 + dz^2),
\label{eq:EH2}
\\
H &=& 1+\frac{1}{e^{-\lambda t}}\left(\frac{M_1}{\sqrt{x^2+(z- a)^2}}+\frac{M_2}{\sqrt{x^2+(z+ a)^2}}\right),
\\
V^{-1} &=&\frac{a/8}{\sqrt{x^2+(z- a)^2}}+\frac{a/8}{\sqrt{x^2+(z+ a)^2}}.
\end{eqnarray}
Similar to the above discussion,
we plot the coordinate values of event horizon
$\tau:= e^{-\lambda t},~x,~z$ in Fig.\ref{fig:EH}.
In Fig.\ref{fig:EHtimeslice}, we plot the 
the coordinate values 
$x,~z$ of event horizon at some typical time slices.
{}From Fig.\ref{fig:EHtimeslice}, we can see 
that two black holes collide at $\tau = -21.100$
and the first contact point is given by $x=0, z=0$.
In this case, $z$-axis is fixed points of $SO(2)$ symmetry in $(x,y)$ space,
but $\tau = -21.100, x=0, z=0$ is not a fixed point of the $U(1)$ isometry generated by $\partial_\psi$.
So the topology of this first contact point in Fig.\ref{fig:EHtimeslice}
is not a point but ${\rm S}^1$
in the whole five-dimensional space-time  since the integral curve of $\partial_\psi$ is ${\rm S}^1$.
This is the specific difference from 5DKT.

\section{Crease Set and Time Slice Dependence}
The intermediate time evolutions of the spatial cross section of
black hole coalescence depends on the choice of time slices in general. 
As discussed in \cite{Siino:1997ix,Siino:1998dq,Ida:2009nd}, 
how black holes coalesce is determined invariantly by the structure
of the crease set of an event horizon and the spatial horizon topology far in the future,
where the crease set is a set of past
end points of null geodesic generators of an event horizon and
the event horizon cannot be smooth 
there.

\begin{figure}[!t]
\begin{center}
\includegraphics[width=0.6\linewidth,clip]{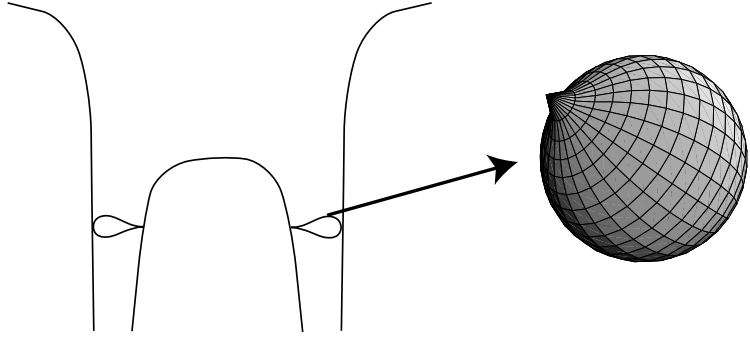}
\end{center}
\caption{
Schematic figure of the event horizon of 5DKT focused on one black hole 
in some time slice before the coalescence occur .
The topology of the spatial cross section of the crease set is a point.
}
\label{fig:pointCS}
\end{figure}

\begin{figure}[!t]
\begin{center}
\includegraphics[width=0.6\linewidth,clip]{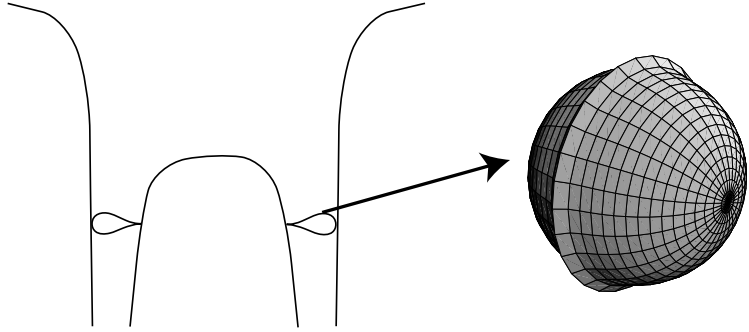}
\end{center}
\caption{
Schematic figure of the event horizon of CBEH  focused on one black hole 
in some time slice before the coalescence occur.
The topology of the spatial cross section of the crease set is ${\rm S}^1$ which correspond to 
a great circle of ${\rm S}^3$.
}
\label{fig:S1CS}
\end{figure}

In 5DKT, 
the crease set is given by the form of
$\tau = \tau(w), x=y=z=0, -a< w < a$. Hence we immediately find that the topology of the crease set is ${\rm R}^1$.
On the other hand, in CBEH, the  crease set is
given by the form of
$\tau = \tau(z), x=y=0, -a < z< a$,
which indicates
${\rm R}^1 \times {\rm S}^1$ in the space-time, 
where ${\rm S}^1$ is generated by $\partial_\psi$.
Clearly, the dimension of crease sets in 5DTK and CBEH are different.
For this reason, choosing 
the $\tau$=constant slice before 
the coalescence occur and focusing on one black hole,
the spatial sections of the event horizons 
can be schematically drawn 
as in Figs.\ref{fig:pointCS}, \ref{fig:S1CS}, respectively.

\begin{figure}[!t]
\begin{center}
\includegraphics[width=0.98\linewidth,clip]{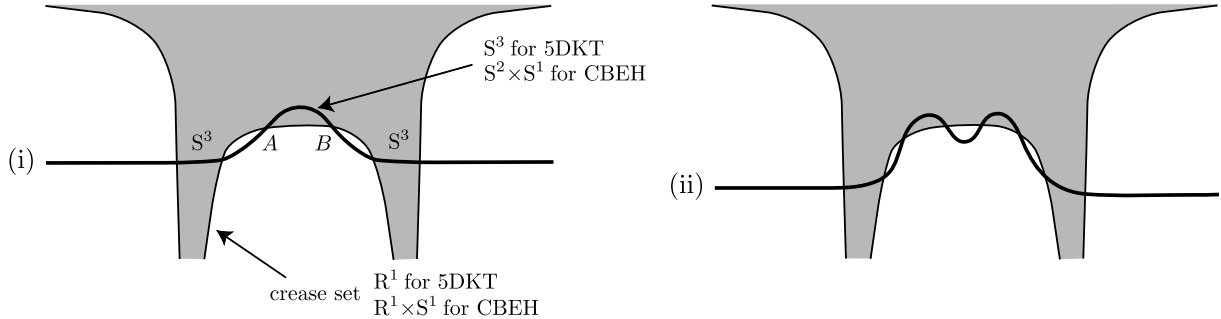}
\end{center}
\caption{Schematic figure of the side view of the event horizon of 
5DKT and CBEH. 
Typical time slices (i) and (ii) which are different from the time slice $\tau$ are depicted.
We assume time slices (i) and (ii) respect $SO(3)$
symmetry for 5DKT and $SO(2)\times U(1)$ symmetry for CBEH.
}
\label{fig:TimeSlice}
\end{figure}
To see the explicit difference between intermediate evolutions, 
we consider another time slice (i)
shown in Fig.\ref{fig:TimeSlice} 
for the both 5DKT and CBEH cases. 
For simplicity, we assume the spatial symmetry is 
the same as $\tau=$const. surface. 
The central part of the intersection of 
the slice (i) and
the event horizon makes a three-dimensional closed surface
in the both cases.

In the case of 5DKT, 
because the points $A$ and $B$ in Fig.\ref{fig:TimeSlice} are fixed points
of the SO(3) rotation in $(x,y,z)$ space,
the middle closed surface in Fig.\ref{fig:TimeSlice} 
is topologically ${\rm S}^3$, i.e., the middle region 
is a black hole with ${\rm S}^3$ horizon.
For more general time slices, 
a number of black holes with ${\rm S}^3$ horizons 
can appear in the 5DKT case  as the time slice (ii)
shown in Fig.\ref{fig:TimeSlice}.

In contrast,
in the case of CBEH,
because the points $A$ and $B$ in Fig.\ref{fig:TimeSlice} are fixed points of the $SO(2)$ rotation but
not fixed points of the $U(1)$ generated by $\partial_\psi$,
the intersection of the slice (i)
and the event horizon is topologically ${\rm S}^1\times {\rm S}^2$.
Then a black ring ${\rm S}^1\times {\rm S}^2$ is formed
in the time slice (i)
during coalescence of two black holes in the CBEH case.
This is due to the differences of the structure of the crease set.

\section{summary and discussion}
In this paper, we have studied how the structures of the event horizons
of five-dimensional coalescing black holes differ
in association with the asymptotic structure. 
We especially focus on the two solutions, i.e. 
Kastor-Traschen solution(5DKT) \cite{Kastor:1992nn,London:1995ib}
in which two black holes with 
${\rm S}^3$ topology coalesce into a 
single black hole with ${\rm S}^3$ topology
and 
coalescing black hole solution on Eguchi-Hanson space(CBEH)\cite{Ishihara:2006ig}
in which two black holes with ${\rm S}^3$ 
topology coalesce into a single black hole with the lens space $L(2;1)$ topology. 
It  came out that, 
if we choose the time slices 
in which the symmetry of the base space is respected, 
the first contact points of 
the coalescing process  compose ${\rm S}^1$ in the CBEH case
unlike the 5DKT case in which the  first contact point is a point 
in the five dimensional spacetime.  
This is the specific difference 
in topology changing process 
${\rm S}^3$ into the lens space $L(2;1)$ in the CBEH case.

The fact the  first
contact point is ${\rm S}^1$ in CBEH can be understood as follows.
Let us consider the coalescence of two spheres centered at each nut of
the Eguchi-Hanson space which is the base space of CBEH
instead of considering the coalescence of black holes.
If the two spheres are chosen to be symmetric
the coalescence would occur at the midpoint between the two nuts.
The topology of the midpoint between the two nuts is ${\rm S}^1$
which corresponds to
equator of ${\rm S}^2$-bolt of the Eguchi-Hanson space in which
two nuts are on the north and the south pole respectively.

We have also investigated the structure of crease sets of the event horizons. 
We have shown that the topologies of crease sets 
are ${\rm R}^1$ for 5DKT and ${\rm R}^1 \times {\rm S}^1$ for CBEH. 
Since the dimensions of the crease sets are different  from each other,
it is expected
that we can see the difference of intermediate evolutions explicitly if
we  adopt a different timeslice.
In fact, 
we can find the time slice in which
black ring ${\rm S}^1\times {\rm S}^2$ is formed  in a certain period
in the case of CBEH unlike the 5DKT.

\section*{Acknowledgments}
We would like to thank Masaru Siino and Ken-ichi Nakao for useful discussions.
This work is supported by the JSPS Grantin-Aid for Scientific Research No. 19540305.
MK is supported by the JSPS Grant-in-Aid for Scientific Research No. 20-7858.
ST is supported by the JSPS under Contract No. 20-10616.

\end{document}